\documentclass[twocolumn,showpacs,preprintnumbers,amsmath,amssymb,showkeys]{revtex4}

\usepackage{graphicx}
\usepackage{dcolumn}
\usepackage{bm}
\usepackage{color}

\begin{document}

\preprint{Superlattices and Microstructures, Volume 30, Issue 4, October 2001, Pages 215-219}

\title{Broadening processes in GaAs $\delta$-doped quantum wire superlattices}

\author{T. Ferrus \footnote{Present address : Hitachi Cambridge Laboratory, J. J. Thomson Avenue, CB3 0HE, Cambridge, United Kingdom}}
\email{taf25@cam.ac.uk}
\author{B. Goutiers}
\author{J. Galibert}
\author{F. Michelini}

\affiliation {Institut National des Sciences appliqu\'ees, 135 Avenue de Rangueil, 31077 Toulouse, France}

\date{\today}

\begin{abstract}

We use both Quantum Hall and Shubnikov de Haas experiments at high magnetic field and low temperature to analyse broadening processes of Landau levels in a $\delta$-doped 2D quantum well superlattice and a 1D quantum wire superlattice generated from the first one by controlled dislocation slips. We deduce first the origin of the broadening from the damping factor in the Shubnikov de Haas curves in various configurations of the magnetic field and the measured current for both kinds of superlattice. Then, we write a general formula for the resistivity in the Quantum Hall effect introducing a dephasing factor we link to the process of localization.

\end{abstract}

\maketitle

Full article available at doi:10.1006/spmi.2001.1007

\end{document}